\documentclass{article}
\usepackage{amsmath,amssymb}
\usepackage[noblocks]{authblk}
\usepackage{epstopdf}
\usepackage{cite}
\usepackage{siunitx}
\usepackage{color}
\usepackage{ulem}
\usepackage{here}
\usepackage{appendix}
\usepackage{graphicx}
\usepackage{wrapfig}
\usepackage{bm}
\usepackage{comment}
\usepackage{sidecap}
\usepackage[top=28truemm,bottom=28truemm]{geometry}

\begin{document}
\title{
Arranging the order of passengers on the boarding bridge to reduce the boarding time for single-aisle aircraft
}
\author[1]{Sakurako Tanida}
\author[1,2,3]{Katsuhiro Nishinari}
\affil[1]{Research Center for Advanced Science and Technology The University of Tokyo, 4-6-1 Komaba, Meguro-ku, Tokyo, Japan}
\affil[2]{Department of Aeronautics and Astronautics, Graduate School of Engineering, The University of Tokyo, 7-3-1 Hongo, Bunkyo-ku, Tokyo 113-8656, Japan}
\affil[3]{Mobility Innovation Collaborative Research Organization, The University of Tokyo, 5-1-5 Kashiwanoha, Kashiwa-shi, Chiba 277-0882, Japan}
\affil[ ]{\textit {u-tanida@g.ecc.u-tokyo.ac.jp}}
\date{\today}
\maketitle

\begin{abstract}
Reducing the aircraft boarding time is a common problem not only for airlines, but also for passengers and airports. Group boarding is a popular boarding strategy that separates the passengers into several groups and those groups, which are then called in a certain order. Group boarding can reduce the boarding time compared with that in random order boarding; however, it is insufficient in several real scenarios because the passengers are not separated strictly into groups. In this paper, we propose a boarding strategy that arranges the order of the boarding passengers at the boarding gate. Although this approach appears more time-consuming, we show that such a rearrangement can be applied efficiently to the waiting queue of a single-aisle aircraft. We quantitatively demonstrate the boarding times for various patterns of this approach and discuss the mechanism underlying the reduction of boarding time. This strategy is a promising approach to reduce boarding times and can replace the conventional boarding strategy.
\end{abstract}

\section{Introduction}
Reducing aircraft boarding time is beneficial for not only the airlines but also the passengers and airports. For airlines, reducing the boarding time reduces the cost of staying at the airport, which constitutes a majority of turnaround time from arrival to departure in an airport (\cite{Horstmeier2001,VanLandeghem2002,Jaehn2015}). More specifically, shortening the boarding time even by 1 min saves US\$30-250 (\cite{Horstmeier2001,Nyquist2008}). In addition, reducing the turnaround time of aircraft and crew could lead to additional flights. Passengers can use the saved time effectively instead of being stuck in a restless boarding process, and may result in a satisfying flying experience. Airports can increase the runway usage efficiency by reducing the turnaround time of each aircraft owing to the shortened boarding time. For the above reasons, a methodology to reduce the boarding time is sorely required.

Currently, for already assigned seats, the boarding strategy popularly employed by several airlines is group boarding. Group boarding, or call-off system, is a strategy where passengers are separated into groups according to seat numbers, and each group is called in order at the boarding gate (\cite{VanLandeghem2002,VanDenBriel2005,Ferrari2005,Bazargan2007,Nyquist2008,Bachmat2008,Steffen2012,Soolaki2012,Tang2012,Qiang2014,Jaehn2015,Giitsidis2016,Schultz2018}). Various ways of grouping are used, such as back-to-front, outside-in, and reverse-pyramid. In back-to-front grouping, boarding begins at the rear of the aircraft and works forward, whereas it begins at the window side seats and works to the aisle side in outside-in grouping. The reverse-pyramid is a mixture of back-to-aisle and outside-in. Among them, the outside-in and reverse-pyramid methods have been reported to reduce the boarding time compared with random boarding (\cite{VanDenBriel2005,Ferrari2005,Bazargan2007,Nyquist2008,Steffen2012,Tang2012,Qiang2014,Jaehn2015,Giitsidis2016,Schultz2018}). In addition, calling each passenger individually has been shown to dramatically reduce the boarding time (\cite{VanLandeghem2002,Ferrari2005,Steffen2008a,Nyquist2008,Steffen2012,Tang2012,Qiang2014,Milne2014,Jaehn2015,Notomista2016}).

As previous studies have shown, group boarding can shorten the boarding time under ideal conditions; however, it approaches that of random boarding when the passengers are not strictly separated into groups~\cite{Soolaki2012}. Similarly, the boarding time for passengers individually called at the boarding gate is closer to random boarding when they are not aligned in the order of calling (\cite{Steffen2008a}). In a practical situation, a group starts to be called before all the passengers of the former group have completed boarding, leading to the mixing of consecutive groups. The timing of group switching is not always optimized because it is mostly determined by the staff on the spot according to their experience. In addition, passengers cannot always board when their group is called. Therefore, even though group boarding is introduced as a system, the expected effect is not achieved in real-life situations.

One way to avoid this problem is to separate passengers into groups and rearrange their order in each group after they have passed through the boarding gate. We call this strategy "group sorting" in this study. In group sorting strategy, we do not need to consider the passing order of passengers at the boarding gate. Although rearrangement after passing through the boarding gate appears inefficient, it can improve efficiency when there is a long waiting queue on the boarding bridge. A long line frequently appears in reality, depending on the seat configuration of the aircraft, such as single-aisle aircraft. In such cases, it is worth applying the group sorting strategy.

In this study, we numerically simulated the boarding model by rearranging the order of passengers after they passed through the boarding gate. First, we discuss the random boarding of the single-aisle aircraft where the passenger waiting queue of passengers appears on the boarding bridge after they have passed the boarding gate. Next, we show that the boarding time increases in non-ideal situations for group boarding from the window to the aisle side because of the mixing of passengers from different groups. Finally, we demonstrate the boarding times of the group sorting strategy for various number of passengers per group, $m$. We assume that the passengers in each group are rearranged from the back to the front seat number on the boarding bridge. The results show that the group sorting strategy for a not-small or not-large $m$ has similar in speed to the ideal group boarding strategy. To clarify the mechanism underlying the rapid boarding time for a not-small or not-large $m$, we examined the relation of the aisle occupation with the boarding time when $m$ of passengers are rearranged from back to front seat numbers. In the discussion, we demonstrate that the number of groups and the aisle occupancy determine the boarding time in the group sorting strategy.

\section{Problem formulation}
\begin{figure}
	\centering
		\includegraphics[width=0.8\linewidth]{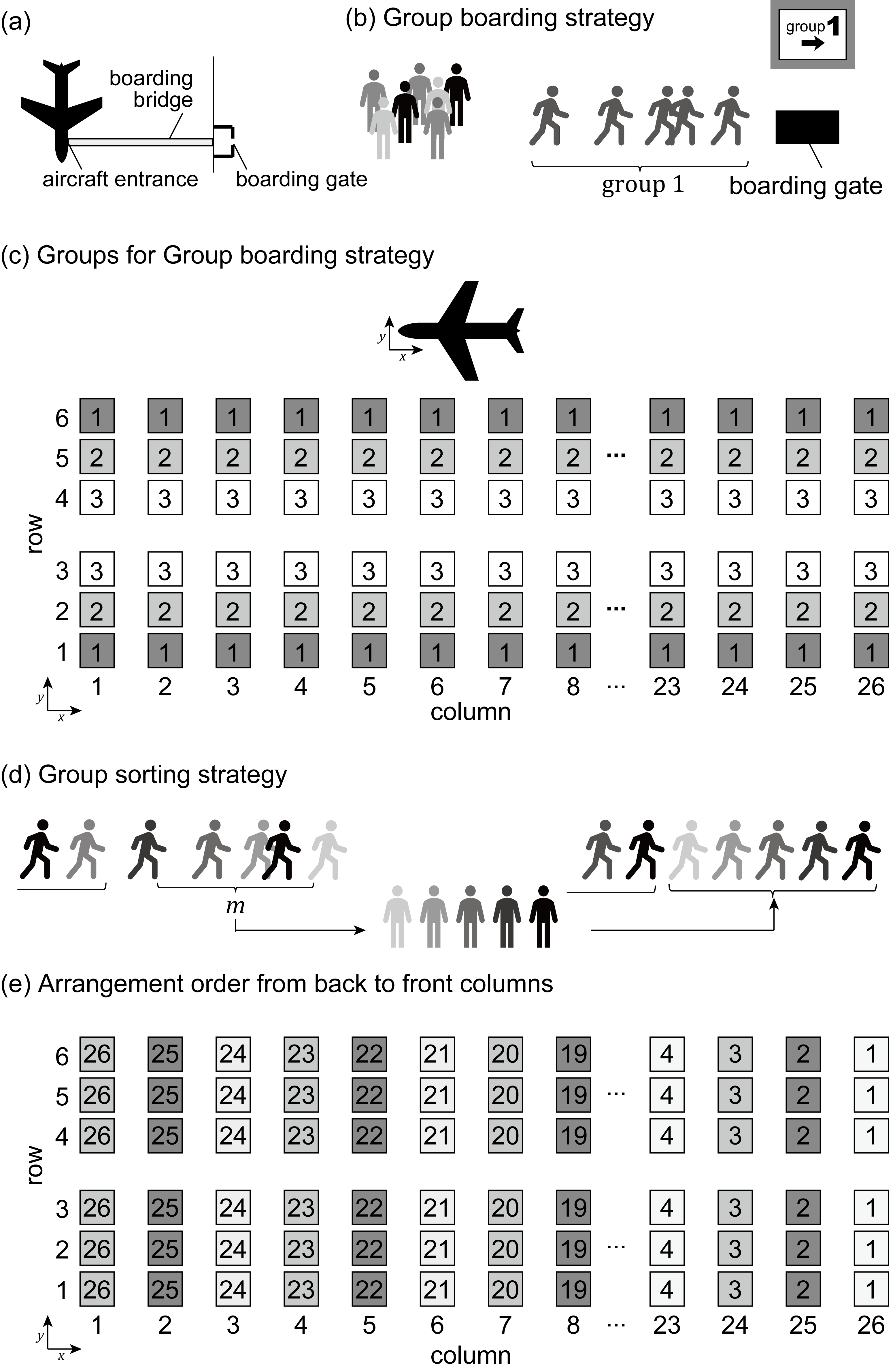}
	\caption{(a) Schematic of the boarding area. (b) Schematic of group boarding. (c) Seat configuration and order of group boarding. (d) Schematic of group sorting strategy. The order of randomly arriving passengers is rearranged at the boarding gate for every $m$ passengers. 
	(b) In group boarding strategy, the order of passengers is rearranged according to the number written on the seat.
	}
	\label{FIG:seats}
\end{figure}

We assumed that passengers pass through the boarding gate, boarding bridge, and aircraft entrance in sequence for boarding an aircraft, as observed at the majority airports worldwide [Fig.\ref{FIG:seats}(a)]. We defined the seat numbers increasing along the body and wing axes of the aircraft as columns and rows, respectively [Fig.\ref{FIG:seats}(c,e)]. We focused on a single-aisle aircraft with six-seat rows. The aisle divides the row into three seats on each side. To simplify the discussion, we ignored the different seat classes and assumed all uniform seats. The total number of seats is $x_{max}=156$, assuming a standard single-aisle aircraft.

We assumed that passengers have tickets with assigned seats. We considered three strategies for deciding the boarding order of the passengers: random boarding, group boarding, and group sorting. In random boarding, passengers gather at the boarding gate in random order. In group boarding, the passengers are separated into three groups, and each group is allowed to board for the window to the aisle row [Fig.\ref{FIG:seats}(b,c)]. In other words, an outside-in type of group boarding is employed. In each group, the passenger column order is random. In group sorting strategy, passengers gathering randomly at the boarding gate are divided into groups of $m$ people [Fig.\ref{FIG:seats}(b)]. The number of passengers in a group follows a discrete normal distribution with a mean of $m$ and a standard deviation of 3. The order of passengers in each group is rearranged from the back to the front columns before they enter the aircraft. In other words, passengers are sorted from the largest to the smallest column number [Fig.\ref{FIG:seats}(c)]. When there is more than one passenger with the same column number in a group of $m$ passengers, the passengers are ordered randomly with respect to the row number. We assumed that the rearrangement had been completed before they went through the boarding bridge.

The boarding dynamics of the passengers were simulated with a pedestrian model based on a cellular automaton programmed in Python (\cite{Fukui1999}). The geometry of the aircraft and boarding bridge was divided into cells of size 0.8 [m] x 0.8 [m], which is roughly the area of a standing person. If a cell is occupied by a passenger, no other passenger cannot enter it. In the simulation, passengers appeared at the boarding bridge at the rate of $\nu$ [/s] in the order of each strategy. They walked on the boarding bridge toward the aircraft entrance. The walking speed was set as 1 cell [/s], or 0.8 [m/s], and the length of the bonding bridge is set to be $L_{bridge}=120$ cells or 96 [m]. The seating area starts immediately after the boarding bridge. A single aisle was defined in the aircraft, with a length of $L_{aisle}=52$ cell or 41.6 [m]. In other words, the maximum possible number of passengers standing in the aisle for a column is $C=2$. Assuming a narrow aisle, we prohibited the passengers from passing other passengers in the aisle and boarding bridge in our simulation. Thus, passengers can move to their forward cell only if the cell is vacant.

All passengers take a time of $T_{bag}$ to place their baggage into an overhead rack. If someone has already occupied the seat closer to the aisle on the same column number, we assumed that it takes another $T_{change}$ time for the passenger to occupy their seat. We set $T_{bag}=10$ [s] and $T_{change}=15$ [s] in this study. Before the completion of the placement of overhead baggage and sitting in the seats, the passenger occupies the aisle cell at the side of their seat and blocks the aisle.

\section{Results}
\begin{figure}
	\centering
		\includegraphics[width=0.9\linewidth]{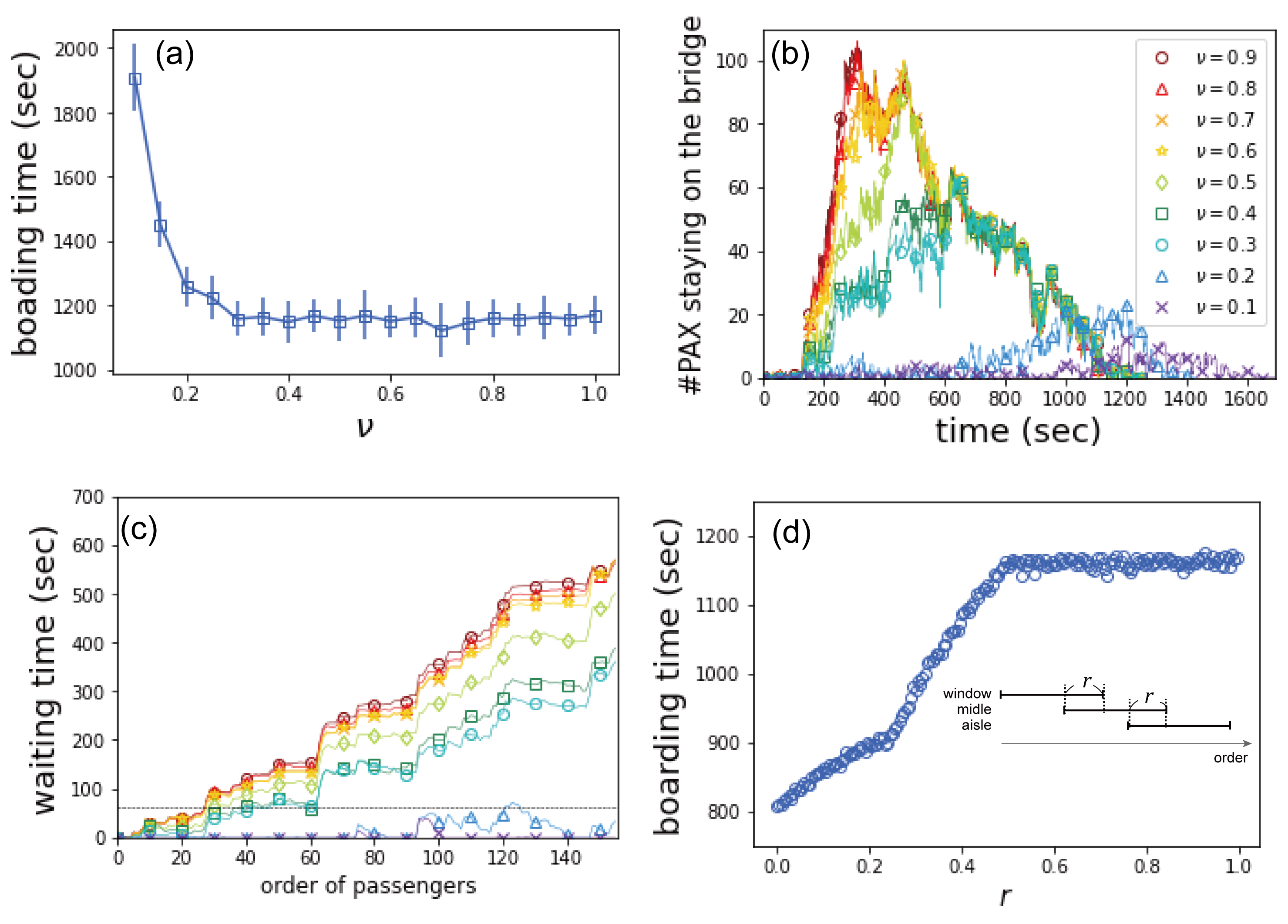}
	\caption{
	(a) Boarding time for random boarding for various inflow $\nu$.
	(b) Time evolution of the number of passengers waiting on the boarding bridge when the passengers board in random order for various $\nu$.
	(c) Total waiting time on the boarding bridge for every passenger for various $\nu$. 
	(d) The boarding time in group boarding when passengers of the next group start boarding only after $r$ passengers in the previous group are remaining. See also the inner image.
	}
	\label{FIG:random}
\end{figure}

First, we considered random boarding for 156 passengers (full capacity). We defined the boarding time as the time since the first passenger enters the boarding gate until all the passengers have occupied their seats. Figure \ref{FIG:random}(a) shows the mean and standard deviation of the boarding time for various inflow $\nu$. For a small $\nu$, the boarding time decreases with increasing $\nu$. Above $v=0.3$, the boarding time is constant at approximately $1.2\times 10^3$ [s]. This suggests that just increasing the inflow, for example, by increase the number of boarding gates, improves the boarding time only up to a limit.

To investigate the waiting queue on the boarding bridge, we examine the time evolution of the number of passengers stopping on the boarding bridge for a certain random boarding order for various $\nu$ [Figure~\ref{FIG:random}(b)]. 
The length of the queue is almost always zero for $\nu=0.1$; however, it displays a measurable value above $\nu=0.2$. Generally, the time evolution is separated into three phases: the number of passengers stopping on the boarding bridge is zero until approximately $t=150$ [s]; it starts to increase from approximately  $t=150$ [s]; it then decreases after a certain time. In the initial phase, all passengers can enter the aircraft without stopping. In the second phase, the passengers form a waiting queue on the boarding bridge, which increases with time. In the third phase, the waiting queue gets solved with time.
Figure\ref{FIG:random}(c) shows the time each passenger waits on the boarding bridge in case of random boarding order for various $\nu$. For $\nu=0.1$ and 0.2, the waiting time for almost all the passengers is less than 1 min [black broken line in Fig.\ref{FIG:random}(c)]. On the other hand, above $\nu=0.2$, the first passengers can enter the aircraft smoothly, whereas subsequent passengers need to stop on the boarding bridge. Notably, if we can rearrange the line of those passengers who need to wait on the boarding bridge, the total boarding time can be reduced in principle. Thus, if a sufficiently large $\nu$ can is provided, rearrangement is worth considering for efficiency.

Next, we investigated the group boarding strategy. As mentioned in the Introduction, group boarding from window to aisle rows is known to be efficient (\cite{VanDenBriel2005,Ferrari2005,Bazargan2007,Nyquist2008,Steffen2012,Tang2012,Qiang2014,Jaehn2015,Giitsidis2016,Schultz2018}). However, in most studies, the mixture among different groups is not considered. To quantify the reduction in efficiency, we examined the boarding time when the passengers in different groups were mixed. Figure~\ref{FIG:random}(d) shows the mean boarding time when the passengers of a subsequent group start boarding when $r$ passengers in the preceding group are still remaining, and the inflow is set as $\nu=0.4$. The boarding time increased with increasing $r$. When $r=0.2$, the boarding time was approximately 100 [s] longer than the ideal case of $r=0$. For $r$ larger than 0.5, the boarding time is equal to that in the random case. This result indicates that the total boarding time will be longer than expected if the passengers in the subsequent groups start boarding before those in the previous group have completed boarding.

\begin{figure*}
	\centering
		\includegraphics[width=0.99\linewidth]{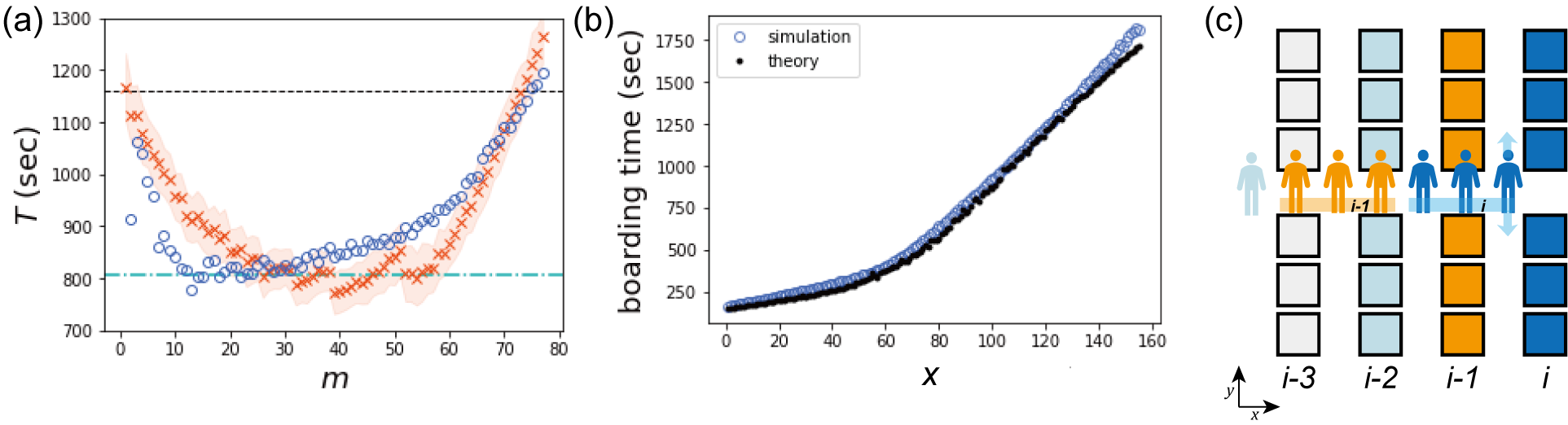}
	\caption{
	(a) The boarding time of group sorting strategy when the mean number of passengers in a group is $m$. The orange cross markers indicate the mean boarding time. The orange shading represents the standard deviation. The black broken line indicates the mean boarding time for random passenger orders. The cyan dash-dot line shows the mean boarding time for an ideal group boarding. Blue circle markers show the estimated value calculated by the equation, which is a function that transforms the value in (b).
	(b) The cyan circles show the boarding time when all $x$ passengers are rearranged in the order shown in Fig.\ref{FIG:random}(e). The black dot markers show the theoretical values calculated using Eq.\ref{eq:est}.
	(c) Schematic of the blocked aisle when two consecutive columns have more than two passengers in the same group.
	}
	\label{FIG:tfinal}
\end{figure*}

Finally, we analyzed the group sorting strategy for full capacity, or $x_{max}=156$. Because the time to rearrange the passenger order depends on the operation method, we calculated the boarding time, $T$, when the passenger order has already been rearranged at the boarding gate, assuming that the rearrangement can be performed without any time loss. The inflow is $\nu=0.4$. The orange markers in Fig.\ref{FIG:tfinal} show the ensemble average of $T$ for various numbers of passengers in a group, $m$. The shaded areas represent the standard deviations. The case of $m=1$ indicates random boarding. The value of $T$ exhibits a downward convex shape against increasing $m$. When $m$ ranges from 20 to 60, the value of $T$ is smaller than for random boarding by 300 [s]. This is almost similar to the boarding time in group boarding where passengers are strictly separated into groups, as shown in Fig.\ref{FIG:random}(d). In other words, it is quicker than most cases of group boarding where the passengers in different groups are mixed.

To clarify the mechanisms of the convex relation of $T$ and $m$, we examined the boarding time when the total number of passengers is $x$ and the order of all passengers is rearranged according to their column number from back to front, $t(x)$. As shown by the blue circles in Fig.\ref{FIG:tfinal}(b), the slope of the boarding time changes at around $x=50$, and the slope of a larger $x$ is steeper than that of a smaller $x$.

The boarding time of $x$ sorted passengers, $t(x)$, corresponds to the boarding time of each group when $m=x$ in the group sorting strategy. The time for the first passenger of $n$ sorted passengers to occupy a seat is $t(1)$ for any $x$. Considering that the number of groups is $x_{max}/m$, the boarding time when $x=m$ can be approximated as follows:
\begin{eqnarray}
	T^{(est)}(m) = \left[ t(m) -t(1) \right]
	        \times \left(\frac{x_{max}}{m}-1\right)+t(1)
	\ ,
    \label{eq:firstapprox}
\end{eqnarray}
where $x_{max}$ denotes the total seat number. The blue circles in Fig.\ref{FIG:tfinal}(a) represent $T^{(est)}$ for various $m$, which can reproduce the tendency of the boarding time during group boarding, as indicated by red crosses in Fig.\ref{FIG:tfinal}(a). This indicates that the slope change of $t(x)$ causes a downward convex for $T$ of group sorting.

Then, why does the slope of $t(x)$ change at approximately $x=50$? Here, we consider the last boarding passenger. If the aisle in front of the last passenger's seat is blocked, they need to wait until already waiting passengers on the aisle are seated. We can roughly estimate that the last passenger is blocked when the number of passengers exceeds approximately 52 because the capacity of persons on the aisle per column is $C=2$. Therefore, the slope of $t(x)$ can change at approximately $x=50$.

To discuss this in more detail, we estimated the total interruption time owing to aisle blocking in the group sorting. We assumed that the blocking occurs at two consecutive columns where more passengers than the aisle capacity, or $C=2$ consecutive passengers, will occupy seats. We label these consecutive columns as $(i-1)$-th and $(i)$-th columns, as shown in Fig.\ref{FIG:tfinal}(c). We color-coded the passengers for the $(i-1)$-th and $(i)$-th columns with orange and dark blue, respectively [Fig.\ref{FIG:tfinal}(c)]. The blocking at the $(i-1)$-th column is solved when $n^{(i)}-C$ passengers have been seated, where $n^{(i)}\in[0,6]$ is the number of passengers in column $i$. Thus, we roughly assumed that  baggage-up and seat-change time for the first $n^{(i)}-C$ passengers of column $i$ is required where $n^{(i)}>C$ and $n^{(i-1)}>C$ are satisfied. More specifically, the total blocking time for baggage-up and seat-change is described as follows:
\begin{eqnarray}
	t^{(block)} = \tau N^{(block)} = \tau\sum_{i \in \Lambda} (n^{(i)}-C)
	\ ,
    \label{eq:blocking}
\end{eqnarray}
where $\Lambda$ is a set of columns that satisfies both $n^{(i)}>2$ and $n^{(i-1)}>2$, and $\tau$ is the estimated time of baggage-up and seat-change. $\tau$ is the estimated time for baggage-up and seat-change. While a baggage-up event is programmed for all passengers, a seat-change event depends on the permutation of the passengers' row. The mean number of seat-changes when one, two, three, and four consecutive persons in the same column are 0, 1/5, 1/2, and 1, respectively. Thus, the expected number of seat-changes per person for one, two, three, and four consecutive passengers is 0, 1/10, 1/6, and,1/4, respectively. Assuming that the number of seat-change is the mean of the expected value, 31/240, $\tau$ is estimated as $\tau = 10+31/240\times15 \sim 12$. Considering that the mean walking time of a single passenger in absence of any other passenger is $L_{bridge}+L_{aisle}/2=146$ cells (or 116.8 [m]) and the inflow is $\nu=0.4$ [/s], the total boarding time for $x$ passengers is obtained as follows:
\begin{eqnarray}
	t^{(est)}(x) = \frac{x}{q}+t^{(block)}(x)+L_{bridge}+\frac{L_{aisle}}{2}
	\ .
    \label{eq:est}
\end{eqnarray}
The black dot markers in Fig.\ref{FIG:tfinal}(b) exhibit the value of $t^{(est)}(x)$ when we employ the ensemble average of $N^{(block)(x)}$ acquired by numerical simulation. $t^{(est)}$ roughly represents $t(x)$. This result indicates that an overflowing aisle in consecutive columns induces a waiting queue that blocks the aisle in other columns. This discussion can be applied to group boarding from window to aisle rows. Because each column of each group has no more than two passengers, aisle blocking owing to the aisle overcapacity is prevented.

Although discussing the effect of group sorting strategy on the boarding time is the main theme in this study, the chain of occupation of the aisle is an additional remarkable phenomenon observed in this study. Considering an overflowing aisle in a column as site-bonding, we can explain it as percolation; namely, the chain of the overflowing aisle in a column seems to be the one-dimensional directed percolation with interaction. Applying the percolation theory to development and verification of the effect of grouping and rearrangement of the boarding passengers could open new avenues for optimizing the boarding strategy.

\section{Discussion and Conclusions}
To reduce the boarding time for a single-aisle aircraft, we propose a boarding strategy called group sorting. In our group sorting strategy, we rearrange the order of passengers arriving randomly at the boarding gate, instead of separating them into groups according to their seat position and calling the groups in a certain order. Although such a rearrangement after arrival at the boarding gate appears to be inefficient, we showed that we can rearrange their order on the boarding bridge, or between the boarding gate and aircraft entrance. More specifically, we demonstrated by numerical simulation that a single-aisle aircraft with full capacity has a long waiting queue from the entrance to the boarding gate. In addition, we demonstrated that the group boarding strategy, which is commonly employed in several airlines, is not always efficient because of the frequent mixing of different groups in practical situations.

Next, we numerically simulated the boarding time for group sorting strategy for various number of passengers per group, $m$. The boarding time exhibits a downward convex against an increasing value of $m$, and is quicker than in random boarding by approximately 300 [s] at most. To investigate the mechanism that shortens the boarding time of the group sorting strategy, we examined the boarding time when $x$ passengers were sorted from the largest to the smallest in the column number in numerical simulation. We found that it can be reproduced using a simple theory that considers aisle blocking, where the neighboring columns' passengers are waiting in the aisle.

These results suggest that group sorting strategy can be a time-effective boarding strategy to replace conventional group boarding. However, the operational method of the group sorting strategy has to be considered, because the rearrangement of passengers should be preferably completed during the period the passengers are waiting between the boarding gate and the airplane entrance. In addition, the operation has to be passenger-friendly for their easy understanding and also for the staff to perform it simply; the application of the group sorting strategy should be designed to fit the configuration of each airport.

\section*{Acknowledgment}
This work was conducted under the Cooperative Research Project Program with ANA HOLDINGS INC.
Part of this work was supported by JST [grant number JPMJMI20D1]. 

%

\end{document}